\documentstyle[aps]{revtex}
\begin{document}
\twocolumn[
\title
{Can Large $u, d$ Quark Masses be Inferred\\
 from the $D_s \rightarrow n \pi$ decay rate?}
\author{J. Milana}
\address
{Department of Physics, University of Maryland\\
College Park, Maryland 20742, USA}

\author{S. Nussinov}
\address
{Department of Physics, University of Maryland,
College Park, Maryland 20742\\
and Physics Department, Tel--Aviv University,
Ramat--Aviv, Tel--Aviv, ISRAEL}
\date{DOE/ER/40762--063, UMPP \#95--011, July 24, 1995.}
\maketitle

\begin{abstract}\widetext
The $D_s \rightarrow n \pi$ measured branching could lead to information on
$u, d$ quark masses if it is dominated by the (helicity suppressed)
$D_s \rightarrow u \bar d$ annihilation channel.  The data would then suggest
anomalously high $m_u \approx m_d \approx 30$MeV masses.  We find
however that $D_s \rightarrow u {\bar d} g$ can explain
$\sum Br(D_s \rightarrow n \pi)$ with zero up and down quark masses.  We
also argue that the present $D_s \rightarrow n \pi$ data may not  even
convincingly imply any annihilation process whatsoever.
\end{abstract}
\vglue0.25in

]

\narrowtext
\section{Introduction}
While lepton masses are known with high precision, confinement and
chiral symmetry breaking effects prevent exact determinations of the  bare
 lagrangian masses of quarks.  This is particularly true for the light $u, d$
quarks.  That $m^0_u + m^0_d \neq 0$ is determined from the
pseudoscalar pions, the Nambu--Goldstone bosons.  The latter get their
mass {\it only} from the explicitly chiral symmetry breaking  terms
$m_u^0 \overline{u} u + m_d^0 \overline{d} d$ in the lagrangian.
Still, the basic relation\cite{GMOR}
$(m^0_u + m^0_d) \langle \overline{q} q \rangle = f^2_\pi m^2_\pi$ leaves
some ambiguity due to the unknown vacuum expectation value of the light
quarks,
\footnote{Indeed only ratio of quark masses can be extracted using
chiral perturbation theory.  See Ref. \cite{Leut1} for a recent discussion.}
so that only roughly  can one say that  $m^0_u + m^0_d \approx 10$MeV.

Lepton masses are directly determined from kinematics.
\footnote{{\it e.g.} the recent measurement\cite{china}
 of $m_\tau$ at the $e^+ e^-$ collider at Beijing depends on
the onset of the $\tau^+ \tau^-$ signal in the threshold region.}
In principle, inclusive rates like that of $\tau$ decay,
$\tau \rightarrow \nu_\tau +$ hadrons can be described at high energies
($E \gg \Lambda_{QCD}$) via a quark process, say
$\tau \rightarrow \nu_\tau + \overline{u} + d$ for the $\tau$ decay into
nonstrange hadrons and a $\nu_{\tau}$ lepton.  The effect of non--zero
small $u, d$ quark masses on the decay rate is then only
${\cal O}(\frac{m_u^2, m^2_d}{m_\tau^2})$ and is rather small for
$m_{u,d} \leq 10$MeV.

There is another interesting way in which non--zero lepton masses are
manifested which has to do with chirality conservation in the weak
(or any other gauge) interaction.  A given chirality eigenstate of a fermion
with mass $m$ and energy $E$ coincide with the corresponding helicity
state up to $m/E$ admixture of the opposite helicity state.  A well--known
argument based\cite{Perkins} on conservation of $J_z$,
the component of the angular
momentum along the decay axis, forbids then the decay
$\pi \rightarrow l \overline{\nu}_l$ in the $m_l = m_v = 0$ limit.
Such a decay only proceeds via the $m/E$ ``wrong helicity'' components.
The decay rate therefore contains an ``helicity suppression factor''
$(m_l^2 + m_{v_l}^2)/m^2_\pi$, a feature utilized in the kinematic searches
proposed for finding $\nu$ mixings in $\pi, K$ decays\cite{Schrock}.

Can such helicity suppression due to light quark masses manifest itself in
hadronic physics, and in particular, can such considerations lead to an
independent determination of $m^0_u + m^0_d$?   A qualitative manifestation
is expected in the case of glueball physics.  The lowest glueball state is
expected
to have $J^{PC} = 0^{++}$ and a mass in the $1.5-2.0$GeV range.  The
helicity arguments would then prefer $gb \rightarrow \overline{s} s$
({\it i.e.} to hadronic final states with kaons) to $gb \rightarrow n \pi$
({\it i.e.} n pion final states).
\footnote{The 1770 $0^{++}$ glueball candidate suggested
recently\cite{Weingarten}
indeed most strongly prefers to decay into $K \overline{K}$ contained states.}

It has been recently suggested\cite{Vietnam} that the decay of the
$D_s$ (the lowest lying bound state of the charm and strange quarks
$D_s= (c \overline{s}) 0^{-+}$) into $n$ pions,
$D_s \rightarrow n \pi$, can be used to actually infer the $u$ and $d$ quark
masses.   This would indeed be the case if the decay is dominated by
$D_s \rightarrow u \overline{d}$, Fig. (1).   By analogy with the leptonic
decays,
one would then expect
\begin{equation}
\frac{\sum_n  Br(D_s \rightarrow n \pi)}
{Br(D_s \rightarrow \mu \overline{\nu}_{\mu}} \sim
\frac{Br(D_s \rightarrow u \overline{d})}
{Br(D_s \rightarrow \mu \overline{\nu}_{\mu}} \sim
3 {\cal E}_{QCD} \frac{m_u^2 + m^2_d}{m^2_\mu}
\end{equation}
where a color factor and a moderate QCD enhancement factor,
${\cal E} \approx a_1^2 \approx 1.6$\cite{BSW} were included.   Using
the experimental
values\cite{PDG} $Br(D_s \rightarrow \mu \overline{\nu}_{\mu}) = .6 \%$ and
$\sum_n Br(D_s \rightarrow n \pi) \approx 1.5 \%$ (the sum is actually
dominated by the single mode $D_s \rightarrow \pi^+ \pi^+ \pi^-$) large
$m_u \approx m_d \approx 30$MeV mass values are inferred.  The authors
of Ref. \cite{Vietnam} call attention to the severe discrepancy between
these values and the more commonly accepted result
that  $m^0_u + m^0_d \approx 10$MeV \cite{Leut2}.

In the following we point out that (un(!))fortunately there is no QCD puzzle
here.  First, because the decay rate $D_s \rightarrow n \pi$ is not anomalously
large and can be accounted for by gluon emission.  Second, because the present
data does not convincingly prove the existence of any annihilation channel
with or without gluons and could still be accommodated by standard
``spectator'' decay, $c \overline{s} \rightarrow s + u + \overline{d} +
\overline{s}$,
followed by $s \overline{s}$ annihilation.  Finally, even if the
analysis would have
indicated unambiguously that the masses $m_u, m_d$ required to explain a large
$\Gamma(D_s \rightarrow u \overline{d})$ rate are
$m_u \approx m_d \approx 30$MeV, no true conflict with QCD would
necessarily emerge.  The point is that when explored at different energies
(or momentum transfers) the $u, d$ quarks may display different masses
interpolating between a ``constituent'' quark mass $m^c \approx 300-350$MeV
at low energies to $m^0_u \sim m^0_d \sim 10$MeV at very large energies.
Having $m_u^{eff} \approx m_d^{eff} \approx 30$MeV at energies of $1$GeV
$= \frac12 M_{D_s}$, would not present a genuine inconsistency: it would
merely point to the importance of higher--twist effects.

We now proceed to discuss these points in more detail.

\section{$D_s \rightarrow u + \bar d +  g$ decay rate}

We now present our estimate for the rate $D_s \rightarrow u {\bar d} g$ using
perturbative QCD (pQCD) methods.    Taking that the emitted gluon hadronizes
roughly $2/3$ of the time into nonstrange quark pairs, this mechanism
should provide a rough estimate for the total decay rate into strangeless
hadronic states.    The present calculation should be viewed as complementary
to the earlier work of Bander, Silverman, and Soni\cite{Bander}
which used a nonrelativistic quark model approach.
We will comment at the end on the various approximations
that have entered and why we consider this to be merely an order of
magnitude calculation.  Nevertheless, it will amply demonstrate that
there is no real case to make that the decay $D_s \rightarrow n \pi$ is
anomalously large.   Indeed, it may in fact be smaller than expected.

The decay mechanism herein being considered is shown in Fig. (2), where
the operator
\begin{eqnarray}
O_1 &=& \frac{1}{4}\bar s_\alpha \gamma^\mu (1-\gamma_5) c_\beta \,
\bar d_\beta \gamma_\mu (1-\gamma_5) u_\alpha,\nonumber\\
\end{eqnarray}
 is a color octet interaction.  Its Wilson coefficient
$C_1(\mu) \neq 0$ at scales $\mu < M_W$ due to QCD evolution.\cite{evolve}

In the  pQCD motivated approach we start with the lowest order
Fock component of the $D_s$ meson.  The decay rate then involves a
perturbatively
calculable hard amplitude convoluted with a nonperturbative,
soft physics wavefunction, $\psi_m$, of the $D_s$.
The factorization scheme
advocated by Brodsky and Lepage \cite{BrodLep} is employed, whereby
 the momenta of the quarks are taken as some fraction $x$ of the
total momentum of the parent meson weighted by a soft physics distribution
amplitude $\phi(x)$.   A peaking approximation is used for $\phi_{D_s}$
wherein
\begin{equation}
\phi_{D_s}(x) = {f_{D_s} \over 2\sqrt 3} \delta (x-\epsilon).
\end{equation}
The decay constant of the $D_s$ is $f_{D_s}$ (in the convention $f_\pi =
93$MeV)
and $x$ is the light cone momentum fraction carried by the light quark.
The parameter $\epsilon$ in $\phi_{D_s}(x)$ is related to the difference
in the masses of the $D_s$ meson and $c$-quark,
\begin{equation}
M_{D_s} = m_c + \Lambda_{D_s}
\label{massofB}
\end{equation}
whereby $\epsilon = \Lambda_{D_s} /M_{D_s}$.  Note that in a pQCD approach
the mass of the strange quark, being a soft--physics momentum scale
($m_s < \Lambda_{QCD}$),  is taken to be zero.  The $SU_f(3)$ violating effects
 differentiating between the $D_s^+$ and the $D^+$ is absorbed in the parameter
$\epsilon$ for each of the two mesons.

In the present context, the factorization scheme is
augmented by the viability of an $\epsilon$ expansion for the decay amplitude.
 Only those terms in the decay amplitude
that are leading in  an $\epsilon$ expansion are kept.  This is
crucial not only because  higher--order terms in the expansion are
formally higher--twist,  but because gauge invariance is otherwise
lost.\cite{doublepeng}
Thus, only gluon emission from the strange quark is included since it contains
a $1/\epsilon$ arising from the strange quark's propagator.  To go to
higher--order
in the $\epsilon$ expansion (by {\it e.g.} including gluon emission off of the
charm quark) would necessitate going beyond the leading logarithm analysis
being used for $C_1(\mu)$ in order to regain gauge invariance.

The amplitude for the decay $D_s \rightarrow u {\bar d} g$ is given by
\begin{eqnarray}
{\cal M} &= \frac{G}{ q \cdot P_{D_s}}&
\left( \xi_\mu q \cdot P_{D_s} +
\imath \epsilon_{\alpha \beta \gamma \mu} P_{D_s}^\alpha \xi^\beta q^\gamma
\right)
\nonumber\\
&&\times{\bar u} (p_1) \gamma^\mu ( 1 - \gamma^5 ) \nu (p_2),
\end{eqnarray}
where $G$ is given by
\begin{equation}
G = \frac{1}{6\epsilon}  G_F C_1(\mu) V_{cs} V_{ud} g_s(\mu) f_{D_s},
\end{equation}
$\xi$ is the polarization vector of the gluon and in which all
color indices have been suppressed.
The decay rate is then given by
\begin{equation}
\Gamma(D_s \rightarrow u {\bar d} g) = \frac{5}{648 \pi^2}
\frac{\alpha_s\mu) C^2_1(\mu)}{\pi^2 \epsilon^2}
(V_{cs} V_{ud})^2 G_F^2 f^2_{D_s} M^3_{D_s}.
\end{equation}
The decay constant of the $D_s$ is obtained from the decay $D_s
\rightarrow \mu \nu_{\mu}$,
whereby we get the relation that
\begin{equation}
\Gamma(D_s \rightarrow u {\bar d} g) = \frac{\alpha_s(\mu)
C^2_1(\mu)}{\epsilon^2}
\frac{5}{2 \pi} \left(\frac{M_{D_s}}{9 m_\mu}\right)^2
\Gamma(D_s \rightarrow \mu \nu_{\mu}).
\label{rates}
\end{equation}

Using the measured branching fraction\cite{PDG}
$Br(D_s \rightarrow \mu \nu_{\mu}) = ( 5.9 \pm 2.2 ) \times 10^{-3}$,
taking $\epsilon \approx 1/4, \Lambda_{QCD} = .2 GeV$, $\mu = M_{D_s}$,
and including a factor $2/3$ for the hadronization of the gluon into nonstrange
modes, we obtain for our estimate for $D_s \rightarrow n \pi$ that
\begin{equation}
Br(D_s \rightarrow n \pi) (\mu^2 = M^2_{D_s} ) = (1.2 \pm .5 )\%,
\end{equation}
 in roughly good accord with the data.

Clearly a number of approximations have been made to obtain this result.
There is for example  considerable dependence on the scale $\mu$ in the above,
leading to roughly a tripling of the predicted branching fraction
if one takes $\mu = M_{D_s}/2$.
It is also not unreasonable to expect sizable corrections to the
$\epsilon$ expansion,
which, although found to work quite well\cite{systematics} in the case of the
two
body decays of the $B$ meson ($\epsilon_B \leq .1$),  may in fact be
marginal in the
present case.     Factors of two or so corrections are hence certainly quite
likely.
Nevertheless this estimate, as well as that to be found in Ref.
\cite{Bander} obtained
by different means, indicates that at the moment
there is no glaring problem in the decay rate for $D_s \rightarrow n \pi$.

\section{Possible generation of $D_s \rightarrow n \pi$ via Spectator diagrams}

The spectator diagram, Fig. (3), driven by the operator ${\cal O}_2$ in
$H_{eff}$,
\begin{equation}
O_2 = \frac{1}{4}\bar s_\alpha \gamma^\mu (1-\gamma_5) c_\alpha \,
\bar d_\beta \gamma_\mu (1-\gamma_5) u_\beta,\label{oper02}
\end{equation}
is the leading decay mechanism of the charm quark ($C_2 \approx 1.25$) and
in the presence of the spectator $\bar s$ quark found in the $D_s$ meson,
leads to a 4 quark state.  Note that the fact that the $\bar d, u$ quark system
(as well as separately the $s \bar s$ pair) form a color singlet,
distinguishes these diagrams from the annihilation mechanisms
of Section (II).  These features are nicely reflected in the final hadronic
systems
containing mesons with an $s \bar s$ component:\cite{PDG}
\begin{eqnarray}
Br(D_s \rightarrow \phi \pi) &=& 3.5\%,
Br(D_s \rightarrow \phi \rho^+) = 6.5\%,\nonumber\\
Br(D_s \rightarrow \eta \rho^+) &=& 10.0\%,
Br(D_s \rightarrow \eta^\prime \rho^+) = 12.0\%,
\nonumber\\
Br(D_s \rightarrow \eta^\prime \pi^+) &=& 4.7\%
\end{eqnarray}
or strange meson pairs (thru color rearrangement in final state by soft
gluon exchanges)
\begin{eqnarray}
Br(D_s \rightarrow K \bar K) &=& 3.5\%,
Br(D_s \rightarrow K^* \bar K) = 7.5\%,\nonumber\\
Br(D_s \rightarrow K^* {\bar K}^*) &=& 5.6\%.
\end{eqnarray}
Along with $\sim 20\%$ semileptonic ($e \nu_e X, \mu \nu_\mu X$) branchings,
$\sim 7\%$ purely leptonic ($0.6\% \mu \nu_\mu$ and
$(m_\tau/m_\mu)^2(1 - m_\tau^2/M_{D_s}^2)^2$
times as much $D_s \rightarrow \tau \nu_\tau$,
and some genuine multibody decays, this could roughly account for the
full $D_s$ decay.
Thus from this particular viewpoint, there is no need for a strong
$D_s$ annihilation channel.
\footnote{Though such an annihilation channel was originally invoked in
Ref. \cite{Bander}
as in the $D^0$ decay in order to explain the shorter lifetime
$\tau_{D_s} < \tau_{D^0}$.}
It is furthermore conceivable that some  final state interactions will
admix a few percent nonstrange final states and hence account for the observed
$D_s \rightarrow n \pi$ rate even without invoking $D_s$ annihilations.

If the $D_s \rightarrow n \pi$ decays are indeed dominated by $D_s
\rightarrow  u \bar d $,
the pattern of these decay is rather puzzling.
\begin{enumerate}
\item $e^+ e^- \rightarrow $ hadrons proceeding via $e^+ e^-
\rightarrow q \bar q$ produces
at $W = 2$GeV an average multiplicity of $\approx 5$.  Why does the $u
\bar d $ system in
$D_s$ annihilation predominantly hadronize into a $3 \pi$ system?
\item Granting that $D_s \rightarrow u \bar d  \rightarrow \pi^+ \pi^+
\pi^-$ is the dominant
mode, why is the $\rho \pi$ component so small: $Br(D_s \rightarrow
\rho \pi) \leq .3\%$ with
$90\%$ c.l., or $Br(D_s \rightarrow \rho \pi)/Br(D_s \rightarrow 3 \pi) \leq
.2$?
Considering the fact that the $Br(D_s \rightarrow K^+ K^- \pi^+$) is dominated
by the branching into the two body $K^* \bar K$ channel, the lack of the $\rho
\pi$
final state is puzzling.
\footnote{Note however that such a suppression is natural in the
mechanism considered in Section (II),  $D_s \rightarrow  u \bar d g$,
where one expects
a leading hadron formed from each of the three outgoing partons.}
\end{enumerate}

Both these puzzles are readily explained if
$D_s \rightarrow \pi^+ \pi^+ \pi^-$ originates from
$D_s \rightarrow (X^0)_{s \bar s} + \pi^+$ generated via the
standard spectator decay diagram.   The $ (X^0)_{s \bar s}$ should be a
 resonance with a strong $s \bar s$ component which decays into
$\pi^+ \pi^-$ with a substantial rate (but not into $\pi^+ \pi^- \pi^+
\pi^-$!).
These requirements are all satisfied by the $f_{980}$ state.
Amusingly, the original experiment\cite{original} did find evidence
for substantial $D_s \rightarrow f_{980} + \pi^+$ decay, accounting
roughly for $30\%$ of
$D_s \rightarrow \pi^+ \pi^+ \pi^-$.  If the width of the $f_{980}$ is
much larger than the
$\Gamma \approx 50$MeV implied in the fit in this experiment,
$D_s \rightarrow f_{980} + \pi^+$ could conceivably be dominant!
Hopefully this issue will be soon settled by the new Fermilab experiment.

\section{Higher Twist Effects: running, non-perturbative $u, d$ quark masses}

The effects of spontaneous chiral symmetry breaking (S$\chi$SB) are
often parametrized as
giving the original, almost massless $u, d$ quarks substantial
``constituent'' masses,
$m_u^c \approx m_d^c \approx 350$MeV.  Viewing the lagrangian quarks
as bare point--like quanta, the ``constituent quarks'' are then the
corresponding
``quasi--particles'' obtained by attaching a cloud of gluons and $q
\bar q$ pairs.  This cloud
endows the quarks with their constituent mass.

This physical picture suggests that when probed with some momentum transfer or
some
energy $E$ the quarks will exhibit some effective mass $m^{eff}_{u,d}(E)$.
Roughly
speaking, only some ``inner core'' of radius $r \approx 1/E$ of the
``cloud'' around the
constituent quark will be probed in this case and only those chiral
symmetry breaking
effects which are generated on distance scales $\leq r$ will become manifest.

The $m^{eff}_q(E)$ curve interpolates between $m^{eff}_q(E) = m^0_q$
at large energies and $m^c$ at small $E$.  Its actual shape depends on the
QCD dynamics underlying the
S$\chi$SB.  If S$\chi$SB is due to the same mechanism which
generates confinement\cite{casher}, then the constituent quarks
would have hadronic sizes and $m^{eff}_q(E)$ would be expected to
abruptly decrease as $E$ is increased beyond $.3-.5$GeV.
If on the other hand, S$\chi$SB is, to a large extent due to gauge
field configurations of
small distance scales, then the $m^{eff}_q(E)$ curve would be much
flatter.  A  remnant
``constituent'' mass of say $m^{eff}_u \approx m^{eff}_d \approx 30-40$MeV
might then conceivablely linger at $E=\frac12 M_{D_s}$ and manifest in
$D_s \rightarrow u \bar d$ annihilation.  Such high values would
of course have important repercussions,
indicating that nonperturbative effects in the $1-2$GeV
region are much more than previously believed.
\footnote{Thus for example, such higher--twist effects would reduce the
$\Gamma(\tau \rightarrow \nu_{\tau} u \bar d) \approx \Gamma(\tau
\rightarrow \nu_{\tau} +
{\rm hadrons})$ rate by
$\sim 16 m^2_{eff}/m^2_{\tau} = 1 \pm \frac12 \%$ (phase space reduction)
for $m = 30 - 60$MeV.  This exceeds the accuracy of the experimental
determination of
$R_\tau \equiv \Gamma(\tau \rightarrow \nu_{\tau} + {\rm hadrons})/
\Gamma(\tau \rightarrow \nu_{\tau} e {\bar \nu}_e$ and the QCD
perturbative estimates.
For a recent interesting synthesis of several QCD results, including
$\tau$ decay, see
Ref. \cite{EKS}.}
Still no true conflict with the $m^0_{u, d} = 10$MeV determination
follows, although in
view of the previous two sections, this observation is mainly of academic
interest.

\bigskip
{\centerline{ACKNOWLEDGEMENTS}}
One of the authors, S. Nussinov, would like to acknowledge most helpful
discussions with H. J. Lipkin.
 This work was supported in part by DOE Grant DOE-FG02-93ER-40762.

\newpage
\begin{figure}
\vglue 2.25in
\caption{The annihilation diagram for $D_s \rightarrow u \bar d$.
The operator ${\cal O}_2$  is defined in Section (III).
This process, as well as additional gluon emission from the outgoing
light quarks, is helicity suppressed.}
\end{figure}
\begin{figure}
\vglue 2.25in
\caption{The leading contribution for $D_s \rightarrow u {\bar d} g$.
Emission from the charm quark is subleading.  Its inclusion would require
decomposing the operator ${\cal O}_1$ to restore gauge invariance.}
\end{figure}
\begin{figure}
\vglue 2.25in
\caption{The dominant decay mechanism of the $D_s$.}
\end{figure}
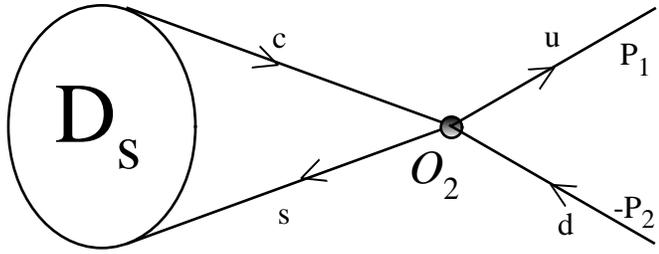
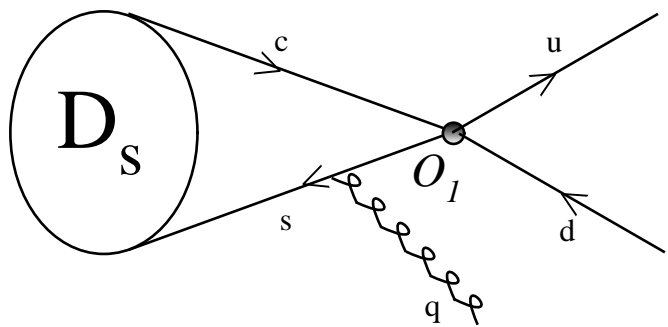
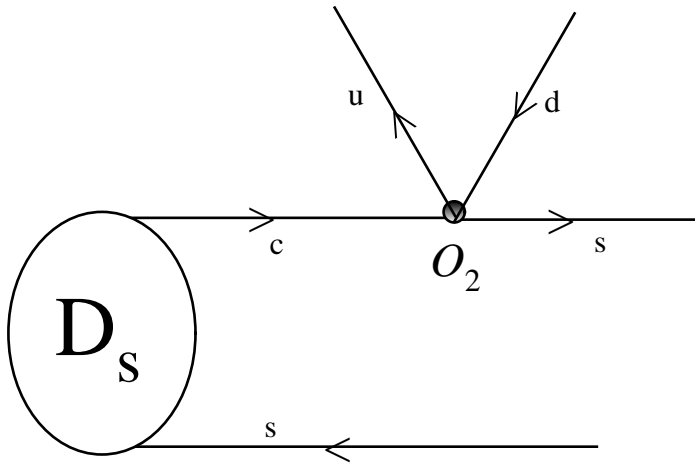
\end{document}